\documentclass[review]{elsarticle}

\usepackage{lineno,hyperref}
\modulolinenumbers[5]

\journal{Renewable Energy}

\begin{document}

\begin{frontmatter}

\title{Resilience and performance of the power grid with high penetration of renewable energy sources: the Balearic Islands as a case study}

\author[mymainaddress,mysecondaryaddress]{Benjam\'{\i}n A. Carreras}
\author[mymainaddress]{Pere Colet}
\author[mysecondaryaddress]{Jos\'e M. Reynolds-Barredo}
\author[mymainaddress]{Dami\`a Gomila \corref{mycorrespondingauthor}}
\cortext[mycorrespondingauthor]{Corresponding author}
\ead{damia@ifisc.uib-csic.es}

\address[mymainaddress]{Instituto de F\'{\i}sica Interdisciplinar y Sistemas Complejos, IFISC (CSIC-UIB). Campus Universitat de les Illes Balears, E-07122 Palma de Mallorca, Spain}
\address[mysecondaryaddress]{Departamento de F\'isica, Universidad Carlos III de Madrid, 28911 Legan\'es, Madrid, Spain}

\begin{abstract}
We analyze the dynamics of the power grid with a high penetration of renewable energy sources using the ORNL-PSERC-Alaska (OPA) model. In particular we consider the power grid of the Balearic Islands with a high share of solar photovoltaic power as a case study. Day-to-day fluctuations of the solar generation and the use of storage are included in the model. Resilience is analyzed through the blackout distribution and performance is measured as the average fraction of the demand covered by solar power generation. We find that with the present consumption patterns and moderate storage, solar generation can replace conventional power plants without compromising reliability up to $30\%$ of the total installed capacity. We also find that using source redundancy it is possible to cover up to $80\%$ or more of the demand with solar plants, while keeping the risk similar to that with full conventional generation. However this requires oversizing the installed solar power to be at least $2.5$ larger than the average demand. The potential of wind energy is also briefly discussed.
\end{abstract}

\begin{keyword}
OPA model; Power transmission grid; Islands decarbonization; Energy transition; 100\% Renewable energy
\end{keyword}

\end{frontmatter}


\section{\label{sec:Intro} Introduction}

The necessity to reduce greenhouse gases emissions to combat climate change and to reduce the dependence of modern society on hydrocarbons is  progressively shifting the energy sector from conventional to renewable energy sources (RES), especially in electricity production. Countries worldwide have road maps to achieve a given percentage of renewable energy in their mix during the next years or decades. This energy transition is particularly pressing on islands, where the electricity production relies mostly on fossil fuels despite their great RES potential. However, their power grid, typically not directly connected to the continental grid, is also relatively small and thus more vulnerable to demand fluctuations and failures in transmission lines and power plants.

Numerous reports inform on the huge potential of RES to provide the necessary energy to run modern societies. For instance, in the case of the Balearic Islands, which we use as a case study in this work, it is estimated that to cover 100\% of the electrical needs with photovoltaic energy, it would be necessary to occupy less than 2\% of the total land \cite{llibre_energies}. These studies are typically based on the mean energy production and consumption depending on different factors, such as mean solar irradiation for solar plants, average wind on certain locations for wind farms, land use, etc. The problems arising from integrating a high share of time variable RES in the existing power grid is often not fully analyzed, although all studies agree on the need to include storage of energy. Indeed integrating high ratios of solar and wind energy in the electric power grid is a challenging task \cite{Sims2011}. 

In this work we address this challenge by focusing on two aspects: On one hand the possible degradation of the grid resilience due an increasing number of blackouts. On the other, the RES performance measured as the fraction of the demand they cover. The evaluation of the resilience and performance is carried out adapting the ORNL-PSERC-Alaska (OPA) model to the Balearic Islands power grid for different scenarios with a high ratio of RES penetration. The OPA model \cite{Carreras2004,dobson2007complex,mei2011} has been widely used in evaluating properties and vulnerabilities of electric power transmission grids. It was validated using several grid models and data of the Western Electricity Coordinating Council (WECC) electrical transmission system \cite{Carreras2013,Carreras2019}.

This paper is structured as follows: In Section \ref{OPA} we introduce the OPA model with the Balearic Islands grid, and the modifications we have made to incorporate RES. In Section \ref{locvsdistr} we find the optimal number and location of the solar power plants to maximize the efficiency and minimize the risk, and discuss the benefits of distributed versus localized generation. Next, in Section \ref{results} we present our main results, and in Section \ref{sec:discussion} we discuss their implications for the transition to an electric systems with high penetration of RES. Finally, in Section \ref{Conclusions} we summarize and give some concluding remarks.

\section{OPA model including RES}
\label{OPA}

The OPA model represents the loads and generators of an electric grid using a standard DC power flow approximation. OPA models the electric grid time evolution in two timescales: a fast timescale of cascading blackouts and a slow timescale that gives the grid evolution over the years. In the fast timescale, the OPA model simulates blackouts through a process of cascading outages of transmission lines. In the slow time evolution, we take an average daily peak load as representative of the daily loading and assume a $0.005\%$ daily increase of the demand (about $2\%$ a year), plus some daily random variations.  

Every day power is dispatched in order to cover the demand giving preference to the RES power plants over conventional ones. Also every day each transmission line has a failure probability rate $p_0$ due to a random event. This outage is a potential trigger of a cascade: in the event of a failure, power is redispatched using the remaining available lines. If, as result of dispatch, a line is overloaded it can fail instantaneously with a probability $p_{1}$. Line connecting nodes $i$ and $j$ is considered overloaded when the ratio $M_{ij}$ of the power transmitted to the power flow limit reaches the value $0.9$.
Power is dispatched again and again until no more line failures are produced. The final solution may have some load shed, $L_{\rm S}$. If the ratio of $L_{\rm S}$ to the total power demand, $P_{\rm D}$ is larger than $10^{-3}$ the event is considered a blackout. The overall stress of the power grid state is measured by $\langle M \rangle=\langle \frac{1} {N} \sum_{i,j} M_{i,j}\rangle$, where $\langle \rangle$ stands for average over the time duration of a simulation, $100000$ days for the cases considered in this work, and $N$ is the number of lines.

When a blackout occurs, the lines that have been overloaded or outaged  during the blackout have their line flow limits increased slightly by a factor $\mu>1$. That is, the parts of the system involved in the last blackout are upgraded. Throughout this paper we take $p_0=0.0001$ days$^{-1}$, $p_1=0.05$, $\mu=1.04$. 

The generation capacity margin $\Delta P_G$ is the difference between the total generation capacity $P_G$ and the power demand $P_D$, normalized to the power demand: $\Delta P_G= (P_G-P_D)/P_D$. The generation capacity is increased every year when its anual average value reaches a predetermined critical value, $\Delta P_G^c$, by upgrading all generators a fixed percentage (4\%). In this way the system copes with the continuous increase in demand. In this work, except for sub-section \ref{more_margin}, we will take $\Delta P_G^c=0.4$. 

Altogether, the OPA model simulates the power grid operating always close to its critical capacity, identifying which are the most probable causes of failure in the system over the years.

\subsection{Balearic Island electric power transmission network}

\begin{figure}
\includegraphics[width=0.75\textwidth]{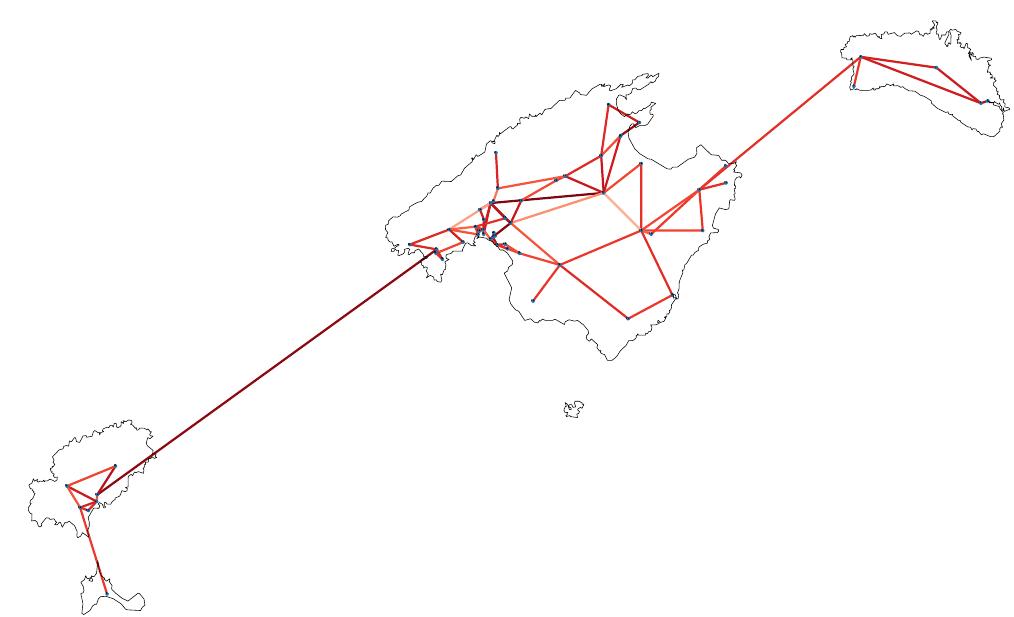}
\caption{\label{fig:network} Network model of the Balearic transmission power grid including the four islands.}
\end{figure}

We consider a model of the Balearic electric grid that includes Menorca, Mallorca, Ibiza and Formentera islands (Fig. \ref{fig:network}). The grid structure and the location of power plants and substations have been obtained from the web site of Red El\'ectrica de Espa\~na \cite{REE}. It comprises $62$ nodes interconnected by $89$ lines. 
There is also a high voltage DC cable connecting the islands with the mainland grid which in this study is replaced by an equivalent additional generation capacity in the conventional power plants. To determine the OPA parameters some information is needed on the performance of the network. At the moment of writing this paper we had not been able to obtain official information on blackouts and their cascading propagation. We resorted to the press for information on the largest blackouts. While this information may be limited and not totally reliable it gives an estimation of the number of large blackouts in the islands. We have estimated the values of the parameters $p_0$ and $p_1$ given above in such a way that the probability of large blackouts (those affecting over 10\% of the population) corresponds to the number of blackouts reported by the press in recent years.

\begin{figure}
\includegraphics[width=0.5\textwidth]{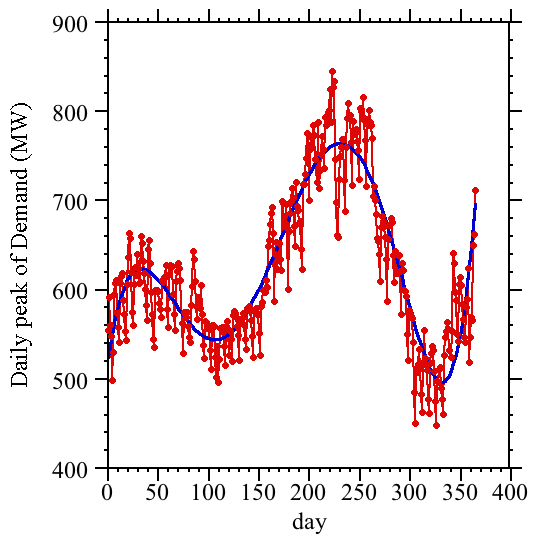}
\caption{\label{fig:demand} Evolution of the daily peak of power demand in Mallorca in the year 2014}
\end{figure}

The daily evolution of the total power demand in the Balearic Islands has been obtained from Red El\'ectrica de Espa\~na \cite{REEdemand}. In particular we consider the daily peak of the demand whose variation along 2014 is shown in Fig. \ref{fig:demand}. As for the spatial distribution of the demand, we use use distribution reported some years ago in Ref. \cite{master} scaled to the actual daily peak.

\subsection{RES power plants: power fluctuations and storage}
\label{Sec:RES}

\begin{figure}
\includegraphics[width=0.5\textwidth]{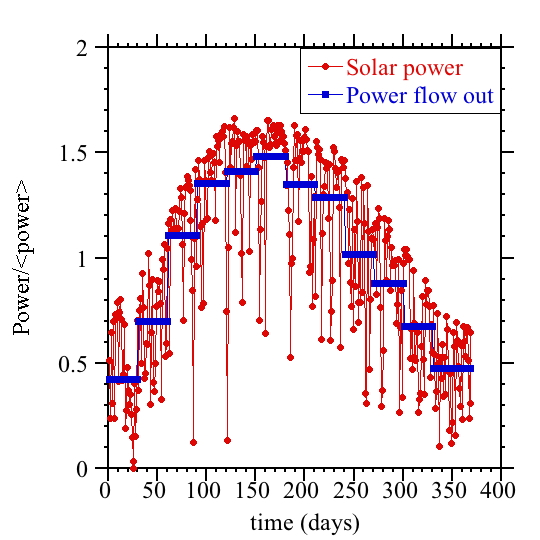}
\caption{\label{fig:solarpower} Daily average power in a photovoltaic solar plant in Menorca (red) and the calculated power flow out month to month of a model plant including storage (blue).}
\end{figure}

The incorporation of renewable energy plants in the OPA model has been discussed in Ref. \cite{Carreras2020}. There it was done for wind generation in the north of California. Here we will follow the same method for solar plants. We base the model on data from a roof-top solar plant on the Consell Insular de Menorca (CIME). The data includes the power production every 15 minutes over nearly three years. Here we will consider the daily average as well as the daily variability as shown in Fig. \ref{fig:solarpower} for the period of one year.

As shown in Fig. \ref{fig:solarpower}, the solar power is highly variable from day-to-day, therefore we assume that solar power plants have also storage capacity in order to guarantee a constant power supply over a certain period. More precisely we consider here that solar power plants aim at delivering a constant level of power each month, in a way that an electric company may contract a fixed power production during this period. This period of time is arbitrary and it can be changed within a certain range without changing the main results to be presented in the following sections. We calculate the power to be delivered, $P_{\rm out}$, by maximizing the performance of the solar plant, i.e. making the most of the available solar energy, and minimizing storage. If $P_{\rm in}(t)$ is the RES power produced every day and $P_{\rm out}(t)$ is the planned power flow out of the plant, we can estimate the energy storage needed to ensure the planned power flow by calculating:

\begin{equation}
 R(t)=\int_0^t \left[P_{\rm in}(t')-P_{\rm out}(t')\right] dt'
\label{R}
\end{equation}

\begin{figure}
\includegraphics[width=0.5\textwidth]{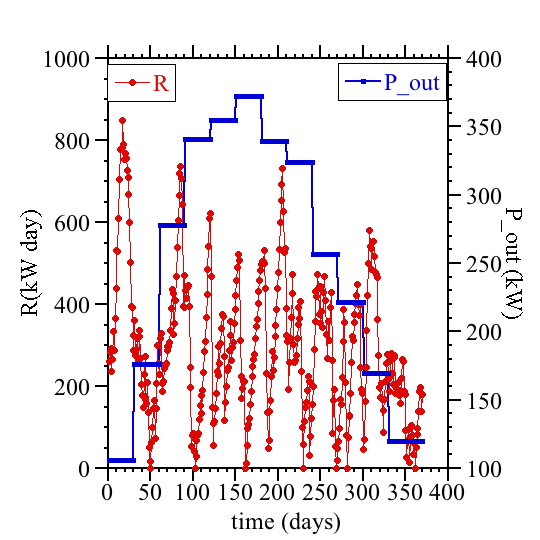}
\caption{\label{fig:Powerflow} Power flow out $P_{\rm out}(t)$ (blue) and energy stored $R(t)$ (red) obtained by minimizing $R_{\rm max}$, Eq.(\ref{R}), for the CIME data.}
\end{figure}

The  maximum value of $R$, $R_{\rm max}$, gives us the needed storage. We calculate the power flow out  $P_{\rm out}(t)$ by minimizing $R_{\rm max}$ with the condition $R (t) > 0$ for any $t$. The result for the CIME data is shown in Fig. \ref{fig:Powerflow}. In this case the minimum necessary capacity for storage turns out to be $R_{\rm max}=850$ kWday while the total output over a year is $\int_0^{365} P_{\rm out} (t) dt \sim 91850$ kWday. Thus the storage capacity is equivalent to less than $1\%$ the annual output, or approximately 3.4 days worth of average energy production. Taking a smaller storage capacity
rapidly increases the probability of under supply, while considering larger storage does not substantially improve performance as, given $P_{\rm out}(t)$, $R_{\rm max}$ has been calculated to make profit of all the possible solar energy production.

\begin{figure}
\includegraphics[width=0.5\textwidth]{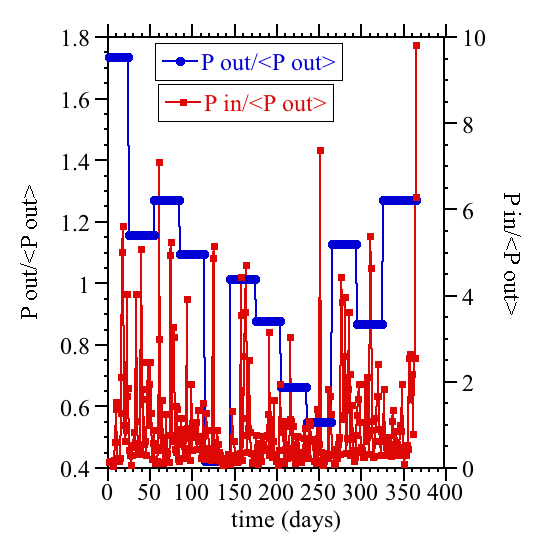}
\caption{\label{fig:Powerflowmodelvent} Daily wind power production $P_{\rm in}(t)$ (red) and the power flow out $P_{\rm out}(t)$ (blue) for wind data from Menorca.}
\end{figure}

We have done a similar analysis using data from the ``Es Mila'' wind power plant in the island of Menorca (Fig. \ref{fig:Powerflowmodelvent}), however because of the large variability of the wind in the Balearic Islands, wind power plants have a low performance and, in large amounts, would induce a high risk for the system.  Therefore we will not include them in this study for now. We will discuss the possibility to include a small fraction of wind power in Section \ref{sec:discussion}.

Next, based on the analysis of the empirical data, we construct a RES power plant model to be included in the OPA code. The power delivered by a plant $P_{\rm g}(t)$ is, whenever possible, equal to $P_{\rm out}(t)$. However, depending on the variable RES generation, $P_{\rm in}(t)$, which we model as a monthly average plus a daily Gaussian random value to simulate daily variations, $P_{\rm g}(t)$ can take different values. If $P_{\rm in}(t)$ is greater than $P_{\rm out}(t)$, then $P_{\rm g}(t)=P_{\rm out}(t)$ and the excess power is accumulated in the storage. If the storage is full the energy is dumped. If $P_{\rm in}(t)$ is smaller than $P_{\rm out}(t)$ power is taken out from the storage, if available, and $P_{\rm g}(t)=P_{\rm out}(t)$. In the case there is not enough stored energy, the delivered power is $P_{\rm g}(t)=P_{\rm in}(t)$.

In this work we consider that the nominal capacity of a RES plant is given by the average of $P_{\rm out}(t)$ over one year, $\langle P_{\rm out}(t)\rangle$. For each plant, the capacity of the storage is taken as $R_{\rm max}=\kappa \langle P_{\rm out}(t)\rangle$ with $\kappa = 3.4$ days, so that it is approximately equivalent to $1\%$ the annual output.

\begin{figure}
\includegraphics[width=0.5\textwidth]{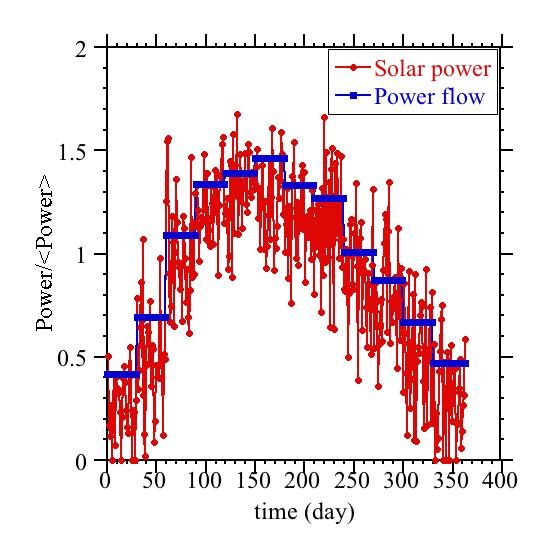}
\caption{\label{fig:Powerflowmodel} Solar power daily production $P_{\rm in}(t)$ (red) and power flow out $P_{\rm out}$ (blue) for the solar plant model in OPA.}
\end{figure}

Fig. \ref{fig:Powerflowmodel} shows the solar power daily production $P_{\rm in} (t)$ and the power flow out $P_{\rm out}(t)$ for the solar plant model. In the figure both $P_{\rm in} (t)$ and $P_{\rm out}(t)$ are normalized to the nominal power, $\langle P_{\rm out}(t)\rangle$. The results of this figure can be compared with the real data shown in Fig. \ref{fig:solarpower}.

Another question that we face when we incorporate solar plants in the OPA model is the degree of correlation in the sun variability at different locations. Since the Balearic islands cover a relatively small territory the correlation is strong for many locations. In this study we consider two uncorrelated sets of solar plants distributed over the network. Solar plants within each set have fully correlated solar insulation. Similarly we consider that wind power plants, to be discussed in Section \ref{sec:discussion}, are distributed in two uncorrelated sets with fully correlated wind fluctuations within each set.

\section{Localized vs distributed generation}
\label{locvsdistr}

In this section we analyze the effect of the number and location of solar power plants on the overall resilience and performance of the system. This is done by considering scenarios with different number $n$ of solar plants. We define the total solar power $P_{\rm S}$ as the sum of the power delivered by the solar plants,
\begin{equation}
 P_{\rm S} (t)= \sum_i^n P_{{\rm g},i} (t).
 \label{eq:P_S}
\end{equation}
Similarly we define the installed solar power as the sum over solar plants of the nominal capacity $\langle P_{{\rm out},i} \rangle$,
\begin{equation}
 P_{\rm SG} = \sum_i^n \langle P_{{\rm out},i} (t) \rangle.
 \label{eq:P_SG}
\end{equation}
The ratio of the installed solar power to the total generation capacity $P_{\rm SG}/P_{\rm G}$ can be considered as a measure of the solar energy penetration in the electric sector and it will be used in what follows to characterize different scenarios.

We perform a Monte Carlo like calculation such that for each scenario with $n$ solar plants we consider $256$ runs with different realizations of the fluctuations and in which solar plants are in each run randomly distributed over the grid.

For each realization the power grid performance is measured as the fraction of the demand covered by solar plants averaged over the time duration of each realization, $\langle P_{\rm S}(t)/P_{\rm D}(t)\rangle$. The risk is calculated as discussed in Ref. \cite{Carreras2014}, and it is a measure of the cost of the blackouts, which depends on both their frequency and size. We use the risk normalized to the case without RES. 

\begin{figure}
\includegraphics[width=0.5\textwidth]{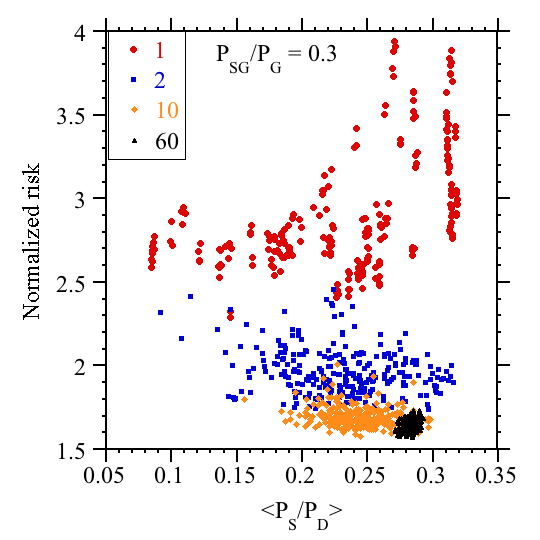}
\caption{\label{fig:Risk} Normalized risk versus the fraction of the demand covered by solar plants considering scenarios in which the installed solar capacity is distributed among $n$ solar plants. For each scenario we plot the results obtained in 256 realizations where solar plants were placed randomly in the grid. Here the installed solar capacity is 30\% the total generation capacity.}
\end{figure}

Fig. \ref{fig:Risk} shows the normalized risk versus the performance measured for each realization of scenarios with a different number of power plants. In all scenarios the installed solar power is $30\%$ of the total generation capacity, $P_{\rm SG}/P_{\rm G}=0.3$, and we consider all the solar plants with the same nominal capacity $\langle P_{{\rm out},i}(t)\rangle=P_{\rm SG}/n$ for $i=1,\dots, n$. Although we have explored many values of $n$, for the sake of simplicity we include in the figure only the scenarios with $1$, $2$, $10$ and $60$ solar plants. We first note that for solar generation concentrated in a single plant, the risk is about three times that of the case without RES and thus grid resilience is seriously compromised. Furthermore there is a large variability in the risk (spread of the different realizations (red points) along the vertical axis) and in the performance (spread along the horizontal axis). 
A noticeable decrease in the risk is observed in going from $1$ to $2$ power plants. This is due to the fact that the two plants are placed one in each set of solar plants with uncorrelated fluctuations. This reduces the probability of having a day without enough power generation, thus reducing the risk of blackouts. Furthermore, risk variability, and to a lesser extend performance variability are also reduced. As we further increase the number of solar plants $n$, the risk and its variability keeps reducing up to $n \approx 10$, beyond which considering a larger number of plants leads to similar results for the risk. Thus, the precise location of the power plants becomes less relevant for grid resilience as we further distribute solar generation beyond $10$ power plants. 
Performance variability also decreases with $n$ but at a slower rate. For $n \geq 60$ solar power plants generate between $27\%$ and $29\%$ of the power regardless of the realization. 

For a given scenario with $n$ solar plants, among all the realizations considered we determine the optimal solar plant distribution as the realization with the lower normalized risk among those with the largest performance $\langle P_{\rm S}(t)/P_{\rm D} \rangle$. This optimal distributions of solar plants will used in the rest of this paper. 

\begin{figure}
\includegraphics[width=0.5\textwidth]{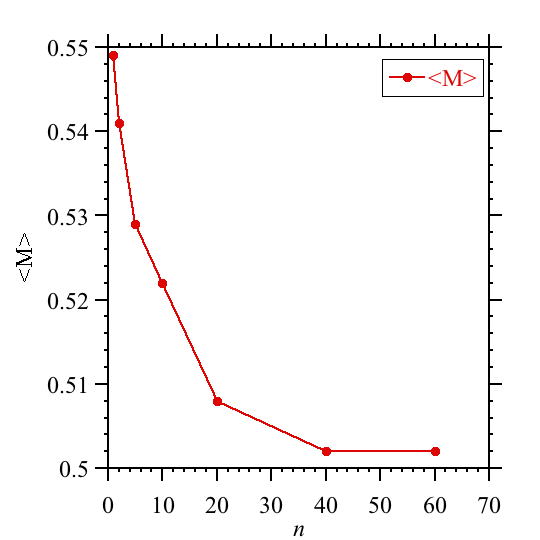}
\caption{\label{fig:M} Grid stress $\langle M \rangle$ for the optimal distributions with $n$ solar plants. Here the installed solar capacity is 30\% the total generation capacity.}
\end{figure}

Typically distributed generation decreases the average current in the lines, reducing the needed transmission capacity \cite{Carreras2011}. This effect can also be noticed here. Increasing $n$ leads to a decrease in the overall stress of the lines as measured by $\langle M \rangle$. Fig. \ref{fig:M} shows the grid stress for the optimal distribution of solar plants with different $n$. While optimal plant distributions for different $n$ perform similarly, distributed generation indeed decreases the average current in the lines as shown in the figure. From $n=1$ to $n=40$ the grid stress decreases about $10\%$ although for larger $n$ it becomes practically independent on the number of solar plants. The benefits of distributed versus centralized generation are not very pronounced here because the Balearic Islands grid is small. In the cases studied in Ref. \cite{Carreras2014} the effect was stronger because larger networks with higher level of interconnection were used.

\section{Replacing conventional generation by RES}
\label{results}

In this Section we consider the optimal distribution for a given number of solar plants and we increase the 
solar penetration ratio $P_{\rm SG}/P_{\rm G}$ from $0$ to $1$ and analyze the behavior of the system. We first do so considering that the total installed power equals the demand plus a margin, which we keep around $40\%$ (subsection \ref{fixed_installed}). Then, in sub-section \ref{more_margin} we will allow for source redundancy, i.e. a much larger amount of solar generation capacity will be installed. The main issue we aim to address in both sub-sections is how far RES penetration can go while maintaining the performance and risk at reasonable levels.

\subsection{Fixed installed power/demand ratio}
\label{fixed_installed}

\begin{figure*}
\includegraphics[width=0.5\textwidth]{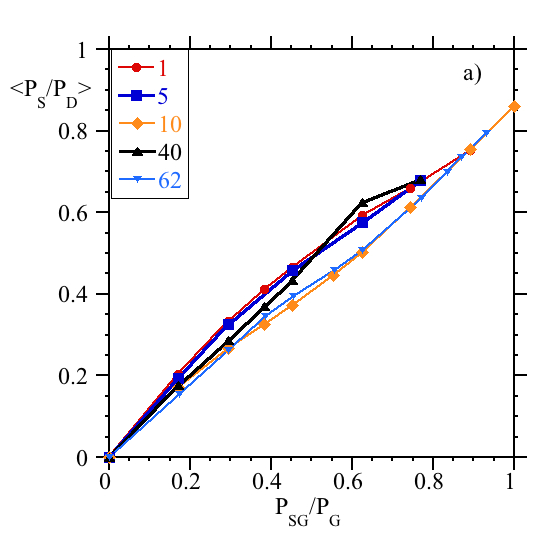}
\includegraphics[width=0.5\textwidth]{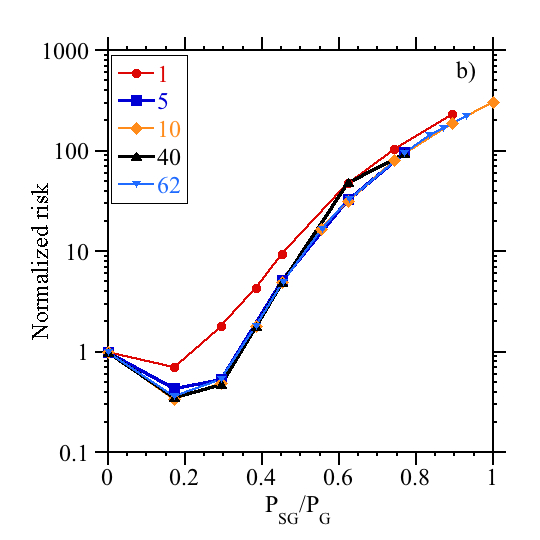}
\caption{\label{fig:fractionSolar} Fraction of the demand covered by solar power (a)) normalized risk (b)) as a function of the fraction of installed solar power for the optimal distribution with different number of solar plants.}
\end{figure*}

In Fig. \ref{fig:fractionSolar} we show the fraction of the demand covered by solar power (panel a)) and the risk (panel b)) as function of the fraction of installed solar power for the optimal distribution with different number of solar plants. The fraction of demand covered by solar generation increases almost linearly with the solar installed capacity up to $P_{SG}/P_{G}\sim 0.4$, point beyond which the growth is sublinear. The risk normalized to the case without RES shows a clear difference between the scenario in which solar generation is concentrated in a single plant and those in which is distributed among $5$ or more plants. The scenario with a single plant has always a higher risk. The reason is two fold. First a single solar plant can not benefit from the uncorrelated fluctuations of the two solar sets we have considered. Second a very powerful power plant in a single node induces higher stress to some power lines, as shown in Fig. \ref{fig:M}. The latter is a general drawback associated to any grid configuration in which generation is very localized. 

\begin{figure}
\includegraphics[width=0.5\textwidth]{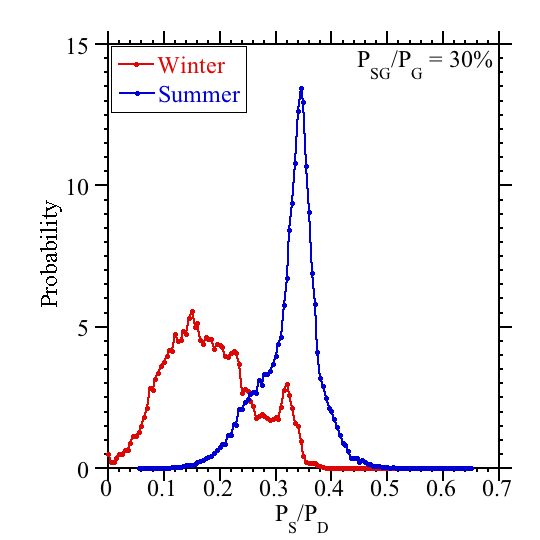}
\caption{\label{fig:winter_summer} Probability distribution function of the fraction of the demand covered by solar plants in winter and summer months. Here the installed solar capacity is 30\% the total generation capacity.}
\end{figure}

We also observe that the general behavior of the risk when the fraction of solar power increases is, initially to decrease, and for fractions larger than $0.3$, to increase exponentially. The initial decrease is due to the benefits of a more distributed generation, while the increase for lager fractions of installed solar power can be attributed to the variability of the solar irradiation, which when the fraction of solar power is large, precludes guaranteeing the supply under all circumstances, inducing blackouts more often. As a matter of fact, the increase is exponential due to an exponential increase in the frequency of blackouts. These blackouts are due insufficient generation to cover the demand. They are not cascading blackouts due to the failure of power lines but they are single step blackouts with lack of generation. This is mainly caused by the low solar generation in winter. We can see in Fig. \ref{fig:winter_summer} the difference in performance between summer and winter for a case with 30\% installed solar power generation. The figure plots the distribution of the fraction of the demand covered by solar plants. During summer month the distribution peaks at $34\%$ and for most part of the time solar plants are producing above 30\% of the demand. In winter the probability distribution broadens significantly and the peak drops to a fraction of the demand around $15\%$.

Naturally as the fraction of installed solar power generation increases the resilience problem becomes more serious, and the lack of power in winter aggravates the situation. The blackouts not only happen in the winter months but also in summer because power storage may be depleted occasionally.

\subsection{Source redundancy}
\label{more_margin}

\begin{figure}
\includegraphics[width=0.5\textwidth]{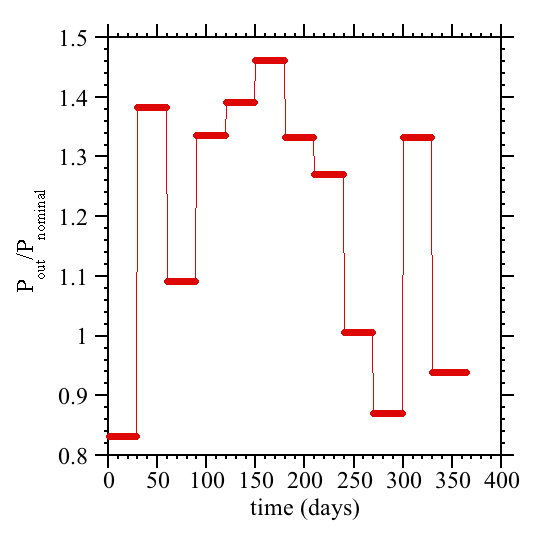}
\caption{\label{fig:powerflow2} Power flow out $P_{\rm out}$ normalized to the nominal capacity with double solar capacity installed in winter months (see text).}
\end{figure}

In the previous sub-section we have shown how, due to the inherent variability of the weather, solar power can not always guarantee the supply. A solution to this problem is installing excess power capacity in order to be able to supply enough energy even during periods of low solar irradiation. 
In this sub-section we consider increasing the installed power in order to be able to fulfill the demand in winter. Such installed capacity is clearly excessive in summer, so we consider a reduced operation during the summer months. We do so in the following way: we double the solar installed capacity in the winter months and use as nominal power, $P_{\rm nominal}$, the one associated to the summer installed capacity. Namely, $P_{\rm nominal}$ is defined as average of $P_{\rm out}(t)$ over a year considering only the capacity installed in summer. 
The extra installed capacity not used in summer could be partially exported to mainland through the cable or stored in the form of hydrogen or using other long term storage technologies. That is a possibility which we do not account for here. If we have double installed solar power in winter than in summer we obtain a flatter power flow out of the solar plants throughout the year as shown in Fig. \ref{fig:powerflow2}. In this figure $P_{\rm out}(t)$ is normalized to $P_{\rm nominal}$. Note that since the nominal capacity is defined considering only summer installed capacity and we are doubling the winter installed capacity, $\langle P_{\rm out}(t) \rangle > P_{\rm nominal}$.

To analyze this situation we take the optimal solar plant distribution for $10$ solar plants (already considered in the previous subsection), double the winter solar installed capacity and consider two different values of the critical generation margin, $\Delta P_G^c = 0.4$, which is the value used so far in this work, and $\Delta P_G^c = 0.6$. The increase in the critical margin ensures a higher excess power. 

\begin{figure*}
\includegraphics[width=0.5\textwidth]{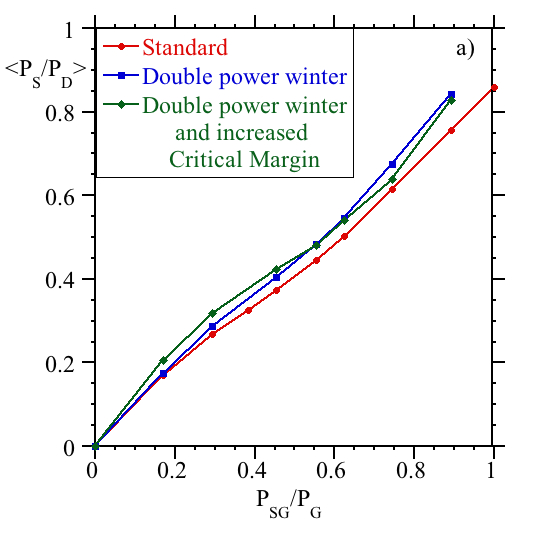}
\includegraphics[width=0.5\textwidth]{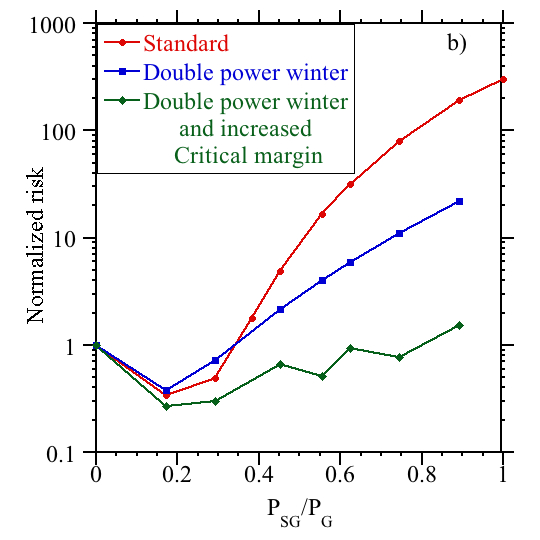}
\caption{\label{fig:fractionSolar2} Fraction of the demand covered by solar power (a)) normalized risk (b)) as a function of the fraction of installed solar power for the optimal solar plant distribution for $n=10$ without source redundancy (red) with doubled installed solar power in winter (blue) and with doubled installed solar power in winter and increased critical margin (green).}
\end{figure*} 

In Fig. \ref{fig:fractionSolar2} we summarize the results and compare them with those of Fig. \ref{fig:fractionSolar} for $n=10$. We observe that the two new configurations with doubled solar installed power in winter have a slightly better performance so that solar power covers a slightly larger fraction of the demand, in particular for large solar penetration. However the most important result is the reduction in the risk. In particular, when we increase the critical margin, we can have a power system operating with a fraction of RES above $80\%$ with the same level of risk as with $100\%$ conventional power.
 
\begin{figure}
\includegraphics[width=0.5\textwidth]{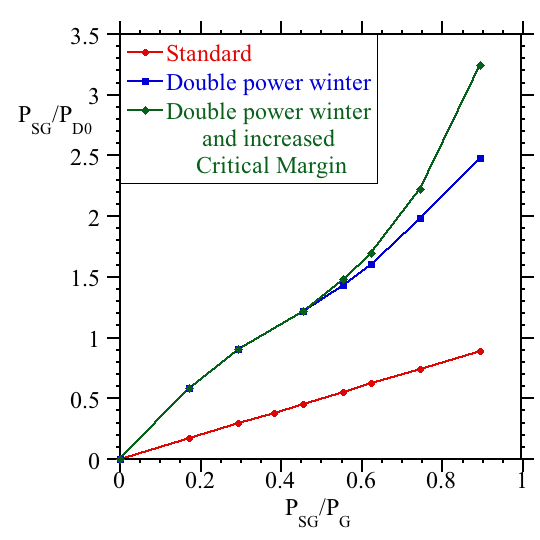}
\caption{\label{fig:powerneeded} Amount of sun power needed to be installed normalized to the annual average demand, $P_{\rm D0}$, for the three cases considered in Fig.~\ref{fig:fractionSolar2} as function of the fraction of installed solar capacity.}
\end{figure} 
 
This impressive reduction of the risk comes at a cost. In order to have double installed solar power in winter and cope with the increased critical margin it is necessary to have a solar installed power which is 2.5 times the annual average of the power demand as shown in Fig.~\ref{fig:powerneeded}. We discuss the implications of such source redundancy in the next Section.

\section{Discussion}
\label{sec:discussion}

The Balearic Islands consumed in $2019$ a total of $6106$ GWh of electric energy, $3506$ GWh in the summer months (May-Oct) and $2600$ GWh in winter (Nov-Apr). This corresponds to an average power of 697 MW. Only $4\%$ of this energy was produced by RES in the islands. Achieving then $35\%$ of the electricity produced by RES by $2030$ and $100\%$ by $2050$ as planned in the 
Law of Climate Change and Energetic Transition of the Balearic Islands (LCCiTE) \cite{Llccite} is a serious economic and social challenge. Including the measures to be taken in other energy sectors, this plan ought to reduce greenhouse gasses emissions by $40\%$ and $90\%$ in $2030$ and $2050$ respectively. 

Considering a mean annual solar radiation of $1.569$ MWh/m$^2$, a $10.8\%$ overall efficiency for the solar plants, and taking into account the slope and orientation of the terrain, a surface of approximately $95$ km$^2$ should be covered by solar panels to produce in average the electricity consumed in the Balearic Islands. This area is less than $2\%$ of the total surface of the islands \cite{llibre_energies}. Roof-top solar panels alone would have the potential to produce, in average, $57\%$ of the total electricity. Solar photovoltaic technology has then, in principle, the potential to produce several times the electrical energy needs of the Balearic Islands. 

This average calculation does not account, however, for the intrinsic variability of the solar energy. Neither it does for storage nor non-variable generation capacity needed to keep the electric system functioning reliably and uninterruptedly. In particular, to produce in average 100\% of the annual energy consumption, the installed peak power (approximately 4400 MWp; 3520 MWp at the AC converters output) would be much larger than the instantaneous consumption at any time (the maximum is around 1250 MW in summer). Therefore storage capacity is needed to properly redistribute the excess production in the moments of maximum solar insulation to the periods of low or null insulation. However, storage in general, and battery storage in particular, is very expensive and very large capacities, such as those allowing to store energy from summer to winter are unpractical. A proper estimation of the installed RES power, storage, and backup non-variable generation needs is imperative for a serious planning and a reliable operation of the system with high penetration of variable RES.

In this work we have estimated the average installed solar power, storage capacity and conventional/non variable power backup needed to keep the system running with a reliability and resilience similar to the actual electric system with almost 100\% conventional power, under different scenarios of RES penetration.
We have addressed the problem of energy supply on a daily time scale. Intraday variability has not been considered nor the frequency fluctuations and grid stability at short times scales, from seconds to hours, which would have required including the effects of primary and secondary control in conventional power plants. This issue will be addressed elsewhere.

In Section \ref{Sec:RES} we have estimated the minimum necessary storage $R_{\rm max}$ needed to compensate for the daily solar variability and keep the power flow out of a solar plant constant over a month. We find that a storage equivalent to less than $1\%$ of the total solar energy produced over a year by the plant is principle enough to use all the available solar power. This storage capacity corresponds to the average production of $3.4$ days. Although a few days worth of storage may seem reasonable, if scaled up for the whole Balearic Islands consumption, it becomes $57$ GWh. To compare with, the largest battery storage projects currently under development have capacities from $0.1$ to $1$ GWh. Storing $57$ GWh of energy would be a challenging problem requiring other technologies besides batteries, such as pumped storage hydroelectric plants, compressed air, molten salts, hydrogen, etc. 

Our model predicts that replacing conventional generation power by solar power with the above mentioned storage, only a $30\%$ penetration of RES can be achieved without increasing the risk too much (Figure. \ref{fig:fractionSolar}). The main problem is that with less that $70\%$ of non-variable generation, the energy supply is not guaranteed in winter months. If RES generation were sized to produce in average the total annual consumption, the power generation would be way too large in summer and too small in winter, and the storage that we have sized for the monthly production would be insufficient to redistribute the summer extra energy to the winter months. 

Without the possibility of storing energy for months, the only solution to cover close to $100\%$ of the demand with renewable energy all year round is sizing the installed solar power to supply enough energy in winter. We have done so increasing the installed solar power until achieving a system with over $80\%$ RES penetration, including source redundancy and increasing the critical margin. While this configuration has a low blackout risk, we have had to increase the installed average solar power up to $2.5$ times the average consumption (Figs.~\ref{fig:fractionSolar2} and \ref{fig:powerneeded}). We note that this scenario still has a installed non-variable power equivalent to $32\%$ ($1.6 \times 20\%$, since the capacity margin is $0.6$ in this case) of the average power demand. Such non-variable power plants could be conventional or based on RES such as biomass or hydrogen produced by excess solar power.
In summer months the large excess energy should be exported to the mainland through the cable, stored in long term storage facilities to maximize the economic benefit, or possibly damped. The present cable has a capacity of $400$ MW which would be less that $5\%$ of the peak power. The economic viability of this solution should be then carefully evaluated. The main drawback of solar power in the Balearic Islands is the low average power in winter. Searching for alternative RES with more availability in winter months would significantly reduce the need of source redundancy.

\begin{figure}
\includegraphics[width=0.5\textwidth]{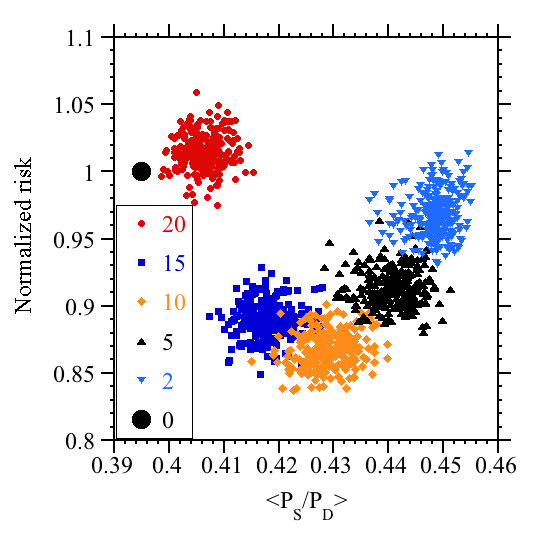}
\caption{\label{fig:wind} Normalized risk versus the fraction of demand covered by RES generation considering scenarios with a different number of solar and wind power plants such that the total number of RES power plants is $62$, one in each node of the power grid. Each symbol corresponds to an scenario with the number of wind plants indicated in the legend. For each scenario we plot the results obtained for $256$ realizations where the nodes with wind plants are chosen randomly. The big black dot is the base case with all RES plants being solar. Here the installed RES capacity is $45\%$ the total generation capacity.}
\end{figure}

Finally, although wind energy fluctuates too much to be a reliable source of energy in the Balearic Island, we have  checked that introducing a small amount, around 15\% of the RES production, is beneficial for the system. We consider a case with $45\%$ RES installed power distributed through $62$ plants of equal power, namely one plant in each node of the power grid. As base case we take all RES plants as solar and we consider scenarios in which a given number of solar plants, randomly chosen, are replaced by wind power plants.  We model wind power plants as the solar plants in Section \ref{Sec:RES} but being $P_{in}$ wind instead of solar power (Fig. \ref{fig:Powerflowmodelvent}).  
Fig. \ref{fig:wind} shows the risk versus the fraction of the demand covered by RES generation for $256$ realizations of each scenario. The risk is normalized to the base case where all RES plants are solar (black thick dot). 
We observe that having up to $20$ wind power plants within the $62$ RES plants increases the RES performance, namely the fraction of the demand covered by RES power. The risk is also reduced, being minimum for the case in which wind accounts for $16\%$ of the RES generation ($10$ power plants). For more than $20$ wind plants out of $62$ the risk increases way beyond the range of the figure. The advantage of including a moderate amount of wind energy comes from the fact that wind fluctuations are in principle uncorrelated with those of the sun, and that there is certain complementary between the availability of wind and sun (Figs. \ref{fig:Powerflowmodelvent} and \ref{fig:Powerflowmodel}b) \cite{Miglietta16}, reducing the overall energy fluctuations. Wind energy has a much higher potential in mainland locations, so RES power can be imported through the cable, benefiting from this complementary beyond what we have discussed in this work.

The considered  scenarios do not include either any management of the demand side. Policies to encourage a reduction of consumption, increase energy efficiency, and use of dynamic demand control as effective storage will be crucial to achieve an electric system with a high penetration of RES in the next decades. 

\section{Conclusion}
\label{Conclusions}

We have analyzed possible scenarios of high RES penetration in the power grid of the Balearic Islands using the OPA model. In this work we consider solar power mainly, as wind is excessively variable in the Balearic Islands to be a reliable power source. We show, however, that a small amount of wind power, up to 15\% of the RES production, have actually a positive effect due to the complementary between sun and wind variability. 

We consider a model for RES power plants that, making use of the least storage possible, takes advantage of as much solar energy as possible, and provides a constant power over one month periods. We find that a 3.4 days worth of energy storage is needed to achieve such performance. Considering RES power plants with such specifications we find that up to 30\% of conventional power can be efficiently replaced by solar power. Beyond this percentage, and keeping the total installed generation capacity equal to the demand plus a $40\%$ margin, the risk of blackouts increases significantly and thus grid resilience is endangered.
One way to overcome this problem is by source redundancy. Doubling the installed power in winter, allows to reach a 80\% RES penetration keeping the risk as low as with full conventional generation. This requires, however, a total installed solar generation capacity $2.5$ times bigger than the average demand. This scenario still requires an installed non-variable power equivalent to $32\%$ of the demand. 
Non-variable power sources can be either conventional, or based on biomass or long term storage from RES, as it could be done for instance using hydrogen as energy vector. Alternative RES with more availability in winter would also reduce the need of source redundancy.

We have therefore shown that an electric system can function with a high penetration of variable RES, but with an important source redundancy and considerable storage needs. The economic and societal cost of implementing such system should be analyzed in detail from an interdisciplinary point of view and, in all cases, measures to increase energy efficiency and reducing consumption will be essential the reduce the need of source redundancy.

\paragraph{Acknowledgmets} The authors DG and PC acknowledge funding
from Ministerio de Ciencia e Innovaci\'on (Spain), Agencia
Estatal de Investigaci\'on (AEI, Spain), and Fondo Europeo
de Desarrollo Regional (FEDER, EU) under grant PACSS
(RTI2018-093732-B-C22) and the Maria de Maeztu program
for Units of Excellence in R\&D (MDM-2017-0711). We thank R. Muñoz Campos and ``Consorci de Residus Urbans i Energia de Menorca'' for sharing data on solar and wind energy production in Menorca, and Red El\'ectrica de Espa\~na for providing data on total energy consumption. B.A.C. and J.M.R.-B. acknowledge access to Uranus, a supercomputer cluster located at Universidad Carlos III de Madrid (Spain).

\end{document}